\documentclass[runningheads]{llncs}
\usepackage[utf8]{inputenc} 
\usepackage[T1]{fontenc}    
\usepackage{lmodern}
\usepackage{multirow}
\usepackage{hyperref}       
\usepackage{url}            
\usepackage{booktabs}       
\usepackage{amsfonts}       
\usepackage{nicefrac}       
\usepackage{microtype}      
\usepackage{graphicx}
\usepackage{float}
\usepackage{lipsum}
\usepackage{caption}
\usepackage{bbold}
\usepackage{subfig}
\usepackage{algorithm2e}
\usepackage{amsmath}
\usepackage{color, soul}

\usepackage{hyperref}
\hypersetup{
    colorlinks=true,
    linkcolor=blue,
    filecolor=magenta,
    urlcolor=cyan,
}
\begin{document}
\title{Adversarial Convolutional Networks with \\ Weak Domain-Transfer for Multi-Sequence \\ Cardiac MR Images Segmentation}
%
%
\author{}
\author{Jingkun Chen\inst{1} \and
Hongwei Li\inst{2} \and
Jianguo Zhang\inst{4,3,1} \and
Bjoern Menze\inst{2}}

\titlerunning{ }
\authorrunning{ }
\titlerunning{Adversarial Convolutional Network with Weak Domain-Transfer}
\authorrunning{J Chen et al.}
%
\institute{}
\institute{University of Dundee, UK
\\
\and
Technical University of Munich, Germany\\
\and
Shenzhen Institute of Artificial Intelligence and Robotics for Society, Shenzhen China\\
\and
Southern University of Science and Technology, ShenZhen China\\
\email{Email: j.f.chen@dundee.ac.uk, hongwei.li@tum.de, jgzhang@ieee.org}}

\maketitle              
\begin{abstract}
Analysis and modeling of the ventricles and myocardium are important in the
diagnostic and treatment of heart diseases. Manual delineation of those tissues in cardiac MR (CMR) scans is laborious and time-consuming. The ambiguity of the boundaries makes the segmentation task rather challenging. Furthermore, the annotations on some modalities such as Late Gadolinium Enhancement (LGE) MRI, are often not available. We propose an end-to-end segmentation framework based on convolutional neural network (CNN) and adversarial learning. A dilated residual U-shape network is used as a segmentor to generate the prediction mask; meanwhile, a CNN is utilized as a discriminator model to judge the segmentation quality.
To leverage the available annotations across modalities per patient, a new loss function named \emph{weak domain-transfer loss} is introduced to the pipeline. The proposed model is evaluated on the public dataset released by the challenge organizer in MICCAI 2019, which consists of 45 sets of multi-sequence CMR images. We demonstrate that the proposed adversarial pipeline outperforms baseline deep-learning methods.

\end{abstract}

\section{Introduction}
Automatic segmentation of the tissues in cardiac magnetic resonance (CMR) images can provide the initial geometric information for surgical guidance \cite{zhuang2016multi}. However, manual delineation of heart structures in CMR scans is laborious and time-consuming.
Late Gadolinium Enhancement (LGE) MR imaging is one of the most effective imaging modalities that can predict heart failure and sudden death \cite{moon2007what}. It enables doctors to visually exam the changes in the myocardium (myo) and confirm the existence of 'cardiomyopathy' and the degree of fibrosis.

There are three main challenges in CMR image segmentation: 1) the large anatomy variations between individuals, and the big diversity of imaging quality in the LGE. For example, due to microvascular occlusion, the contrast agent cannot reach certain areas of the heart, resulting in different enhancements; 2) the ambiguities of boundaries between different cardiac tissues, i.e., the intensity range of the myocardium in LGE CMR overlaps with the surrounding muscle tissue \cite{dataset2}; 3) Despite its clinical importance, LGE slice is much more difficult to annotate than both T2-weight and bSSFP, thus the annotations of LGE CMR are often not accurate or not available. In contrast, the annotations of T2-weight and bSSFP are easier and often available. To tackle these challenges, various methods have been proposed for whole-heart segmentation \cite{zhuang2019evaluation}, ventricles segmentation \cite{petitjean2015right,avendi2016combined}, etc.

In recent years, deep convolutional neural networks (CNNs) \cite{lecun2015deep} have achieved remarkable success in various computer vision tasks \cite{he2016deep,long2015fully} as well as medical image segmentation \cite{unet}.  Generative adversarial networks \cite{gan} as a recent machine learning technique, offers a promising avenue in image synthesis \cite{isola2017image} as well as image segmentation \cite{luc2016semantic}.

We propose a framework to segment ventricles and myocardium from LGE CMR images based on CNNs and adversarial learning, when the annotations of LGE images are rather limited for training.
Our contributions in this work are three folds: 1) we proposed an adversarial segmentation network containing two tailored modules: a segmentation model and a discriminator model, trained and optimized in an end-to-end fashion. The segmentation network generates the predicted masks, and the discriminator network aims to identify the segmentation mask and the ground-truth mask. The segmentation quality is improved in the min-max game. 2) since different modalities share structure information, we introduced a loss function named \emph{weak domain-transfer loss} to leverage information from available modalities with rich annotations;  3) results show that the proposed method outperforms traditional CNN-based method.

\section{Method}
\begin{figure}[t]
  \centering
  \includegraphics[height=6cm,width=12cm]{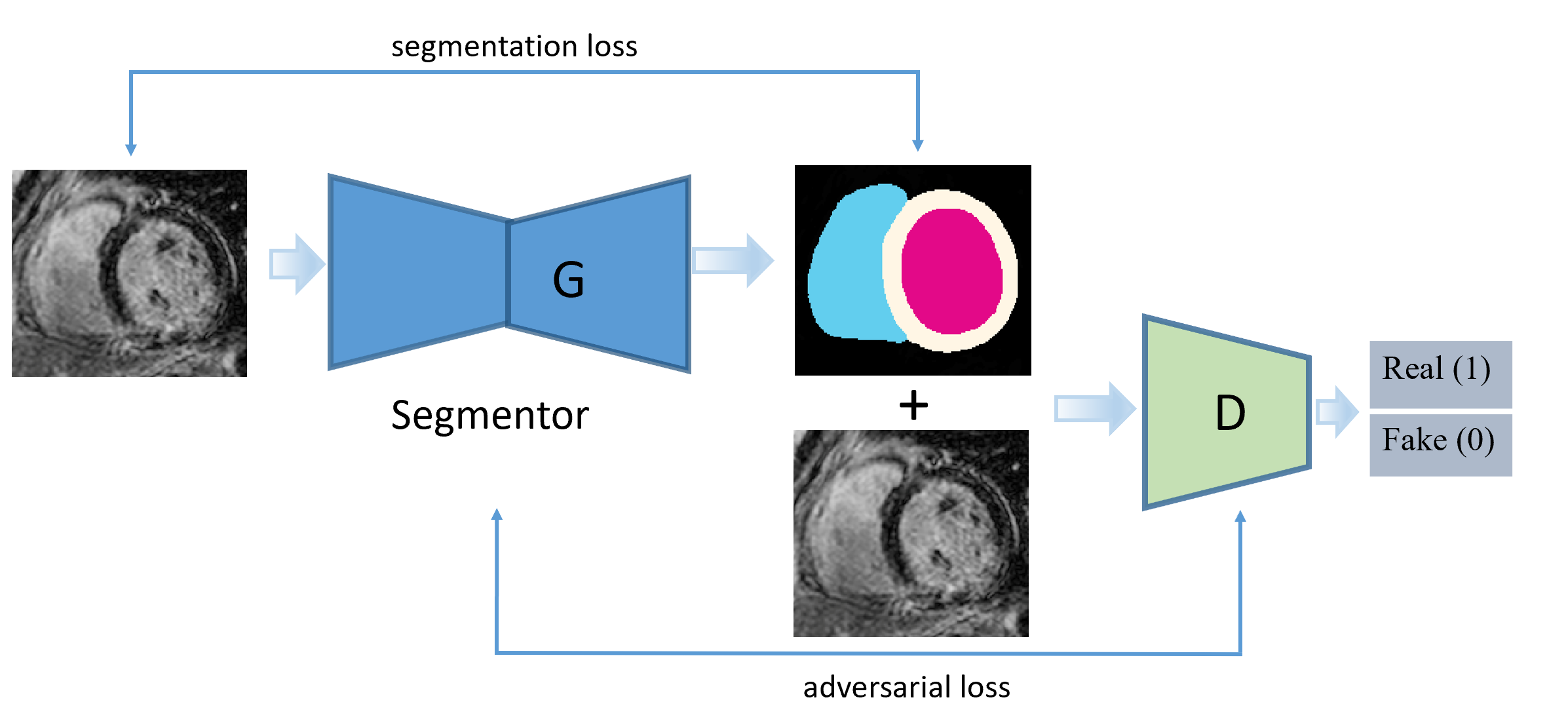}

  \caption{Adversarial segmentation network architecture. It consists of a generator based on a dilated residual U-shape network and a CNN discriminator. The two networks are simultaneously optimized during the process of supervised learning and adversarial learning. Segmentation loss is a combination of individual-domain and domain-transfer loss, while the adversarial loss is a combination of the segmentation loss and the discriminator loss.}
  \vspace{-1.7cm}
  \label{adversarial}
\end{figure}

 Our adversarial segmentation framework consists of a segmentation network and discrimination network. A dilated residual U-shape networks \cite{li2018automatic} is used as a segmentor (i.e. mask generator) \emph{G} and a CNN classifier as a discriminator \emph{D}. \emph{D} is used to ensure that a generated mask being close to its ground truth mask conditioned on the same raw image; the segmentor and the discriminator are updated to improve the performance in an adversarial manner. We also leverage information from other common modalities using a \emph{weak domain-transfer loss}. Figure \ref{adversarial} shows the framework of the proposed method.
\vspace{-0.3cm}
\subsubsection{Data and preprocessing.}
The dataset is provided by the challenge organizers \cite{dataset1} and \cite{dataset2}. It consists of 45 patients, each with three MRI modalities (LGE, T2-weight and bSSFP). It is noted that not all of the modalities come with the annotations of three heart regions (i.e., left ventricles, myocardium, and right ventricles).  Annotations of all the three modalities are provided for patients 1-5; while patients 6-35 have manual annotations of T2-weight and bSSFP. Patients 36-45 have the raw MR scans of three modalities but without any annotations. When constructing the training set, only those MR scans with manual annotations are included. The test data contains the MR scans of LGE from patients 6 to 45, tasked to predict the masks of the three heart regions. Data augmentation is used for robust training. Three geometrical transformations (rotation, shear, zooming) are applied to all of the images and their corresponding masks. For each slice, we also crop a region with a fixed bounding box (224*224), enclosing all the annotated regions but at different locations to capture the shift invariance, resulting in 5 groups of cropped regions of interests. Before training the networks, the intensities of each 2D slice from three modalities are normalized using \emph{z-scores} normalization to calibrate the range of intensities.
\vspace{-4cm}

\subsubsection{Weak Domain Transfer.}
\begin{figure}[H]
	\centering
	\renewcommand{\thesubfigure}{}
	\subfloat{\includegraphics[height=2cm,width=2cm]{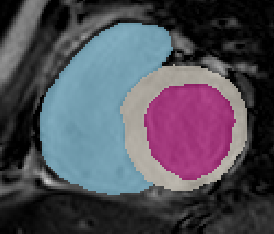}}\renewcommand{\thesubfigure}{} \quad
	\subfloat{\includegraphics[height=2cm,width=2cm]{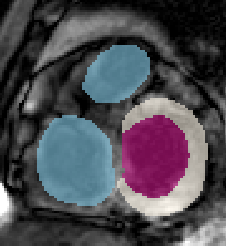}}\renewcommand{\thesubfigure}{} \quad
	\subfloat{\includegraphics[height=2cm,width=2cm]{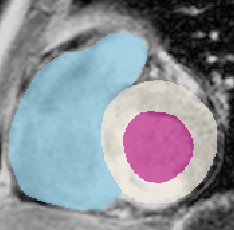}}\renewcommand{\thesubfigure}{} \quad
	\vspace{-0.3cm}
	
	\subfloat{\includegraphics[height=2cm,width=2cm]{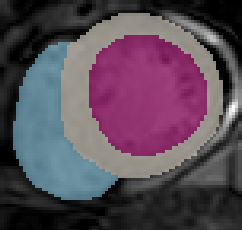}}\renewcommand{\thesubfigure}{} \quad
	\subfloat{\includegraphics[height=2cm,width=2cm]{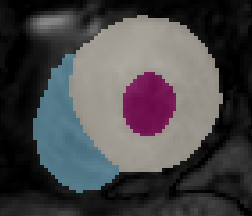}}\renewcommand{\thesubfigure}{} \quad
	\subfloat{\includegraphics[height=2cm,width=2cm]{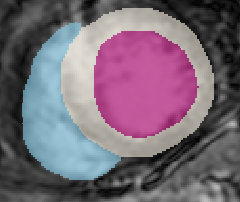}}\renewcommand{\thesubfigure}{} \quad
	\vspace{-0.3cm}
	\caption{From left to right are the images of the bSSFP, T2, LGE modalities from
the same patient, with ground truth masks imposed (best viewed in color).}
	\label{transfer_show}
\end{figure}

Figure 2 shows some sample images with annotation masks of different modalities from the same patient. In Figure \ref{transfer_show}, we can further observe from the annotations that the \emph{bSSFP}, \emph{T2} and \emph{LGE} share some anatomical and structure information; For example, the right ventricle is always surrounded by myocardium, left ventricle is next to myocardium. The annotation masks of the corresponding slices from the three modalities have a certain level of overlap. Based on those observations, we hypothesize that the information from bSSFP and T2 can facilitate the segmentation of LGE. Hence we propose to use the annotation masks on bSSFP and T2 modalities as the \textit{pseudo} masks for the unlabelled LGE modalities.

The masks of bSSFP and T2 scans are transferred to LGE by using a normalized index which identifies the correspondence between axial slices from different modalities. These masks from bSSFP or T2 are directly used as the \emph{pseudo} masks for the corresponding LGE. Specifically, for an axial slice $i$ in bSSFP (or T2) with annotations, its corresponding slice index $j$ in LGE is computed as below:
\begin{equation} \label{equation_1}
\begin{aligned}
j = \lfloor i*\frac{n}{m} \rfloor
\end{aligned}
\end{equation}

\noindent where $\lfloor\cdot\rfloor$ is the floor function. $n$ denotes the number of axial slices of LGE; while $m$ is the number of axial slice in bSSFP (or T2) respectively. Therefore the mask of slice $i$ in bSSFP (or T2) is treated as the pseudo mask of the slice $j$ in LGE.

%




Notably, those masks are \textit{pseudo}, therefore, the domain-transfer loss should be set as a \textit{weaker} one when combined with loss defined on ground truth annotations from expert. We will discuss this further in next section.


It is worth noting that our \textbf{transfer} is different from the \textit{conventional transfer}, which often used a pre-trained model (e.g. on ImageNet), or a knowledge distillation framework of teacher-student learning \cite{hinton2015distilling}. Instead, our \textit{transfer} is built as part of the whole model, specifically tailored for the cross-domain annotation-transfer problem.

\subsubsection{Generator.}
\begin{figure}[H]
  \centering
  \includegraphics[height=6cm,width=12cm]{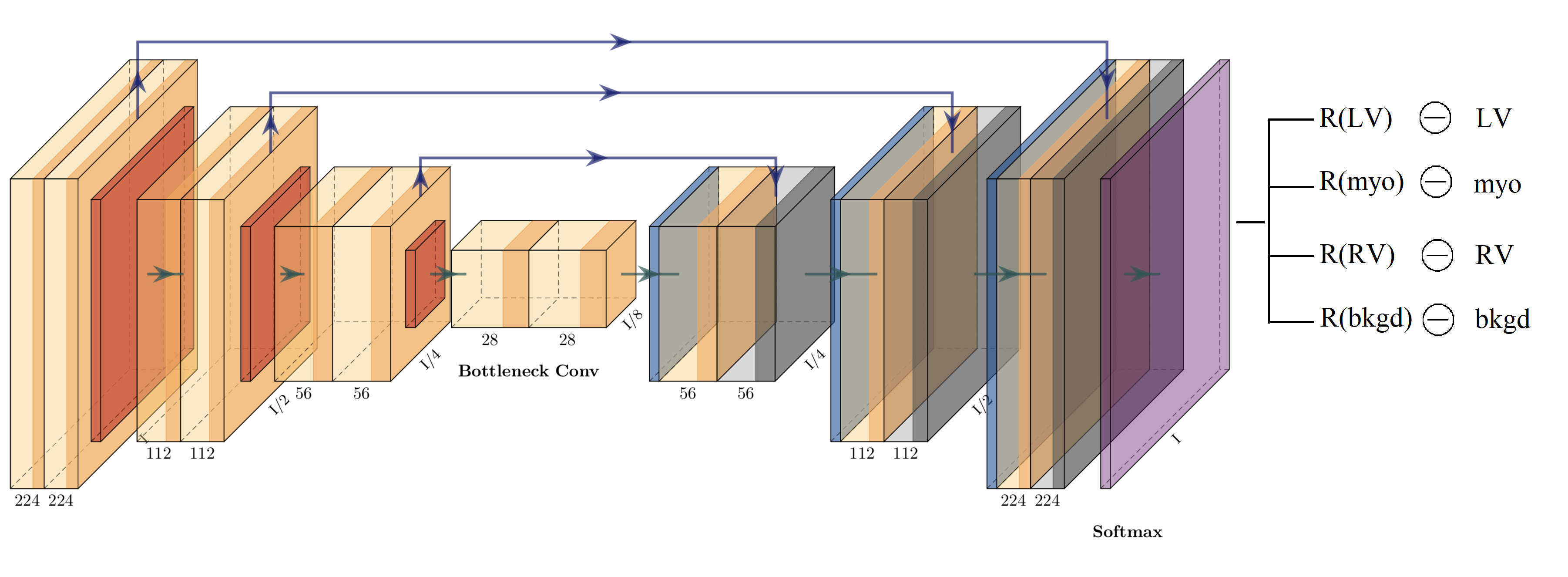}
  \vspace{-0.6cm}
  \caption{Generator network architecture, composed of a downsampling tower and an upsampling tower.}
  \label{generator}
\end{figure}

Figure \ref{generator} shows the overview of the generator model, where a dilated residual U-shape network is tailored and used for the segmentation network. Residual blocks in downsampling and upsampling parts are connected through skip connections. In total the entire network consists of only 0.16 million trainable parameters.

In training a segmentation model, it is aware that cross-entropy loss focuses on individual pixels while Dice loss focuses on the overlap of regions. Thus, a combination of cross-entropy loss and Dice loss is chosen to optimize the network. Images and ground truth masks from the three sequences as well as the transferred masks mentioned above are used. Therefore, the training loss includes two parts: individual-domain loss and domain-transfer loss. Individual-domain loss, denoted as $\mathcal{L}_{ID}$, is the difference between the ground truth mask and prediction while \emph{domain-transfer loss} denoted as $\mathcal{L}_{DT}$, is the difference between transferred masks (pseudo masks) and predicted ones.

Both of $\mathcal{L}_{ID}$ and $\mathcal{L}_{DT}$ consist of a linear combination of the multi-class cross-entropy loss $\mathcal{L}_{ce}$ and the Dice loss $\mathcal{L}_{Dice}$, formulated as:

\begin{equation}
	\mathcal{L}_{ID} = 	\beta_{1}\cdot\mathcal{L}_{ce} + (1-\beta_{1})\cdot \mathcal{L}_{Dice}
\end{equation}
\begin{equation}
	\mathcal{L}_{DT} = 	\beta_{2}\cdot\mathcal{L}_{ce} + (1-\beta_{2})\cdot \mathcal{L}_{Dice}
\end{equation}

The total loss function $\mathcal{L}_{G}$ is formulated as:
\begin{equation}
  \mathcal{L}_{G} = \lambda\cdot\mathcal{L}_{ID} + (1-\lambda)\cdot \mathcal{L}_{DT}
  \label{eq:transfer}
\end{equation}
Notably, the domain-transfer loss leverages the information from bSSFP and T2 modalities. It is worth noting that $\lambda$ in Eq.(\ref{eq:transfer}) is used to control the balance of the transfer; and it is set to 0.9, thus giving a much lower weight of the transfer loss 0.1 which is weak. In our experiments, $\beta_{1}$, $\beta_{2}$ are set to 0.9 after observing the segmentation performance on a validation set.\\
\vspace{-0.7cm}
\subsubsection{Mask Discriminator.} We use a CNN as a discriminator to drive the generator to generate good-quality masks similar to the ground truth ones. The architecture contains several residual blocks with max-pooling layers. The raw images and the masks are spatially concatenated as a multi-channel input to the CNN. A (negative) binary cross-entropy loss $\mathcal{L}_{D}$ is used to train the model, defined as: 

\begin{equation} \label{discriminator}
    \begin{aligned}
    \mathcal{L_D}(\mathcal{S}, \mathcal{T}, D, G) =
    \mathbb{E}_{(x,y)\sim\mathcal{S}}[\log D(y|x)]+
    \mathbb{E}_{(x',y')\sim\mathcal{T}}[\log D(y'|x')]+\\
         \mathbb{E}_{(x,y)\sim\mathcal{S}}[\log (1-D(G(x,y)|x))]+\\
     \mathbb{E}_{(x',y')\sim\mathcal{T}}[\log (1-D(G(x',y')|x'))]\\
    \end{aligned}
\end{equation}

where $\mathcal{S}$ is the set of training data $x$ with ground truth masks $y$, and $\mathcal{T}$ is the set of LGE data $x'$ without masks, but with pseudo masks $y'$.

\subsubsection{Adversarial Training of Generator and Discriminator.}
The objective of the proposed system is to produce appropriate segmentation masks on the target class during the min-max game of the two networks. Firstly we perform a supervised training on \emph{G} using the MR scans with ground truth masks, the objective of G is to generate a good mask to deceive the discriminator network \emph{D}. The goal of \emph{D} is to identify the generated masks from the real masks. We aim to improve the segmentation quality by merging the generated masks with the original images as condition labels and putting them into the discriminator for adversarial learning training. The adversarial model is designed to minimize the adversarial loss which will reverse optimize the generator loss.

Equation \ref{eq:Loss_adv} represents the total loss in the adversarial model. \emph{G} and \emph{D} are simultaneously optimized. 

\begin{equation}
\label{eq:Loss_adv}
	\min_{G}\max_{D}\mathcal{L}_{adv} = \mathcal{L}_{D} + \mathcal{L}_{G}
\end{equation}
\begin{center}
  \textbf{Algorithm 1}: Training procedure of the adversarial model
\end{center}
\begin{algorithm}[H]
\SetAlgoLined

\hrule

{Input: training images X, training masks Y, iteration \emph{j} and \emph{k}, batch size \emph{n}}\\
{Output: Models: Segmentation model \emph{G}, Discriminator \emph{D}}\\
\hrule

i = 0 \\
\While{i < j}{
update \emph{G} by $\mathcal{L}_{IN}$ \\
i = i+1
}
\While{i < k}{
  update D by maximizing $\mathcal{L}_{adv}$ using a mini-batch while keep G fixed\\
  update G by minimizing $\mathcal{L}_{adv}$ using a mini-batch while keep D fixed. \\
  i = i+1 \\
}
\Return G \\
\hrule
\end{algorithm}

\section{Experiment}
\subsubsection{Implementation.} The proposed method is implemented using \emph{Keras} library. The codes are available at \url{https://github.com/jingkunchen/MS-CMR\_miccai\_2019}.  $\alpha$ is set as 0.9 thus, giving the weight of 0.9 for the categorical cross-entropy loss and 0.1 for Dice loss. Learning rate is set to $2\times10^{-4}$, and the learning decay is $1\times10^{-8}$. We use a batch size of 16. For the transfer loss $L_{DT}$, we use the ground truth (whenever available) masks of T2-weight and bSSFP, as the \textit{pseudo} ground truth masks for the corresponding LGE slices. The correspondence between the LGE slices and the T2-weight (or the bSSFP) slices are established based on the simple index normalization along the z-axis of the 3D MRI scans.\footnote{In practice, we find this works well. Ideally, registration could be performed to find the correspondence, which will be investigated further.}. We use Adam optimizer. \\

\subsection{Results}

It is noted that only 5 patients have LGE annotations available, thus we performe a very preliminary experiment to test the proposed method. We held out patients 4 and 5 for testing and the rest for training. Results are reported in Table \ref{tab:result} in terms of Dice score and Hausdorff distance (LV, myo, RV). We further compare three methods: dilated residual U-shape networks with Dice loss (U+D), adversarial model with Dice coefficient loss (U+A+D), adversarial model with Dice coefficient loss and transfer loss (U+A+D+T).
The U-shape networks are specifically designed to segment biomedical images and perform well in myocardial segmentation of bSSFP CMR images \cite{dataset1}. Here we use dilated residual U-shape networks with Dice loss (U+D) as our baseline for a comparison.
It could be observed that adding adversarial training improves the segmentation performance on both the myocardium and right ventricles, but performs worse on left ventricles. The proposed method with transfer loss outperforms both of them with only one exception of the lower Dice score the right ventricle.

\begin{table}[!h]
 \caption{Average Dice and Hausdorff distance on patients 4 and 5}
  \centering
  \begin{tabular}{cc|c}
    \toprule
    \multirow{2}{*}{Method}      & Dice     & Hausdorff Dist.  \\
    \cline{2-3}
        &LV,~myo,~RV & ~LV,~myo,~RV\\
    \midrule
    U-shape network(U+D) & 70.5\%, 50.0\%, 70.0\%  & 13.2, 12.0, 24.6     \\
     Adversarial Model(U+A+D)& 65.1\%, 53.9\%, \textbf{74.7}\% & 38.0, 16.1, 19.4    \\
     Adversarial Transfer(U+A+D+T)& \textbf{76.0}\%, \textbf{59.6}\%, 71.7\%  & \textbf{10.2}, \textbf{12.1}, \textbf{12.9}   \\

    \bottomrule
  \end{tabular}
  \label{tab:result}
\end{table}

\subsubsection{Results on Challenge Test Set.} We submitted the results of the methods of (U+A+D) and (U+A+D+T) on the testing set containing patients 6 to 45 LGE. Table \ref{tab:result_sub1} and \ref{tab:result_sub2} summarize the average and median values of the results returned by the organizers. It could be seen that overall the approach of (U+A+D+T) outperforms (U+A+D), which confirms that promise of the proposed method.

\begin{table}[!h]
 \caption{Average Dice, Jaccard, Surface Distance and Hausdorff distance on patients 6 to 45}
  \centering
  \begin{tabular}{l|c|c|c|c}
    \toprule
    \multirow{2}{*}{Method}      & Dice     & Jaccard   &Surface Dist. & Hausdorff Dist.\\
    \cline{2-5}
     & LV, myo, RV     & LV, myo, RV   &LV, myo, RV & LV, myo, RV\\
    \midrule
     U+A+D& 76.6\%, 42.0\%, 69.5\% & 0.62, 0.27, 0.54  & 5.5, 4.7, 5.5 &22.1, 42.0, 32.7  \\
     U+A+D+T& \textbf{82.4}\%, \textbf{61.0}\%, \textbf{71.0}\%  & \textbf{0.71}, \textbf{0.45}, \textbf{0.57}  & \textbf{3.9}, \textbf{4.0}, \textbf{5.0} & 23.7, \textbf{24.6}, \textbf{23.5} \\

    \bottomrule
  \end{tabular}
  \label{tab:result_sub1}
\end{table}

\begin{table}[!h]
 \caption{Median of Dice, Jaccard, Surface Distance and Hausdorff distance on patients 6 to 45}
  \centering
  \begin{tabular}{c|c|c|c|c}
    \toprule
    \multirow{2}{*}{Method}      & Dice     & Jaccard   &Surface Dist. & Hausdorff Dist.\\
    \cline{2-5}
     & LV,myo,RV     & LV,myo,RV   &LV,myo,RV & LV,myo,RV\\
    \midrule
     U+A+D& 77.8\%, 42.7\%, 71.1\% & 0.63, 0.27, 0.55  & 5.3, 4.3, 5.0 &18.5, 41.2, 28.5  \\
     U+A+D+T& \textbf{82.1}\%, \textbf{60.8}\%, \textbf{72.8}\%  & \textbf{0.70}, \textbf{0.44}, \textbf{0.57}  & \textbf{3.8}, \textbf{3.9}, \textbf{4.6} & \textbf{15.4}, \textbf{19.6}, \textbf{22.8} \\

    \bottomrule
  \end{tabular}
  \label{tab:result_sub2}
\end{table}

\begin{figure}[H]
	\centering
	\renewcommand{\thesubfigure}{}
	\quad\quad \ O \  \quad\quad
	\subfloat{\includegraphics[height=1.3cm,width=1.3cm]{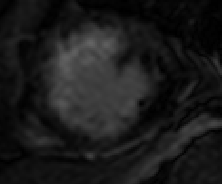}}\renewcommand{\thesubfigure}{} \quad
	\subfloat{\includegraphics[height=1.3cm,width=1.3cm]{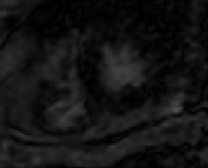}}\renewcommand{\thesubfigure}{} \quad
	\subfloat{\includegraphics[height=1.3cm,width=1.3cm]{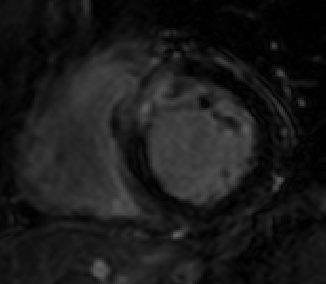}}\renewcommand{\thesubfigure}{} \quad
	\subfloat{\includegraphics[height=1.3cm,width=1.3cm]{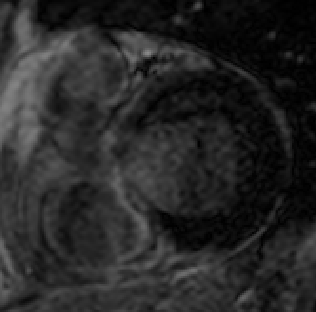}}\renewcommand{\thesubfigure}{} \quad
	\subfloat{\includegraphics[height=1.3cm,width=1.3cm]{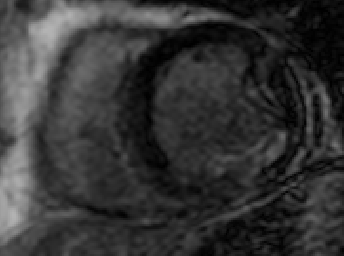}}\renewcommand{\thesubfigure}{} \quad
	\vspace{-0.3cm}
	
	\quad \ U+D\qquad
	\subfloat{\includegraphics[height=1.3cm,width=1.3cm]{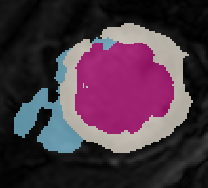}}\renewcommand{\thesubfigure}{} \quad
	\subfloat{\includegraphics[height=1.3cm,width=1.3cm]{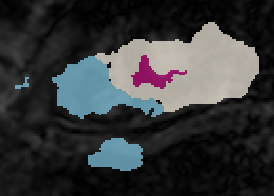}}\renewcommand{\thesubfigure}{} \quad
	\subfloat{\includegraphics[height=1.3cm,width=1.3cm]{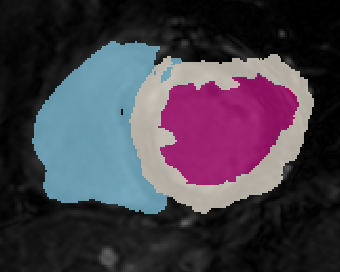}}\renewcommand{\thesubfigure}{} \quad
	\subfloat{\includegraphics[height=1.3cm,width=1.3cm]{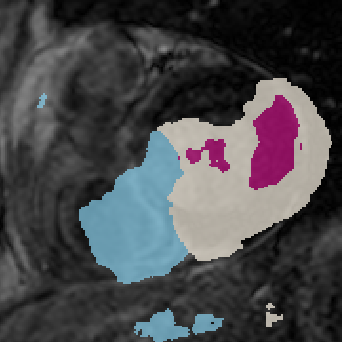}}\renewcommand{\thesubfigure}{} \quad
	\subfloat{\includegraphics[height=1.3cm,width=1.3cm]{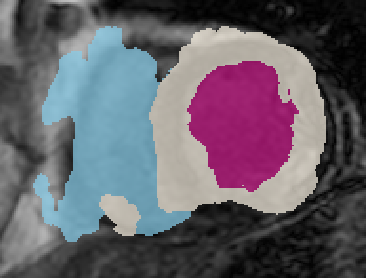}}\renewcommand{\thesubfigure}{} \quad
	\vspace{-0.3cm}
	
	\ \, U+A+D\quad
	\subfloat{\includegraphics[height=1.3cm,width=1.3cm]{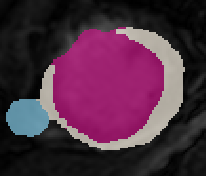}}\renewcommand{\thesubfigure}{} \quad
	\subfloat{\includegraphics[height=1.3cm,width=1.3cm]{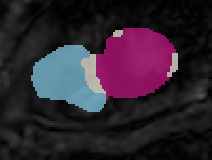}}\renewcommand{\thesubfigure}{} \quad
	\subfloat{\includegraphics[height=1.3cm,width=1.3cm]{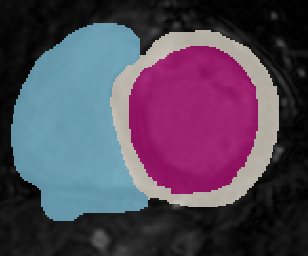}}\renewcommand{\thesubfigure}{} \quad
	\subfloat{\includegraphics[height=1.3cm,width=1.3cm]{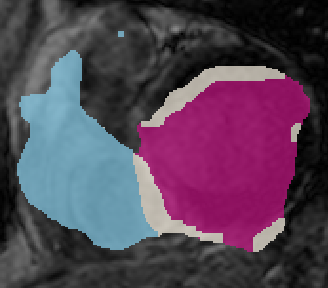}}\renewcommand{\thesubfigure}{} \quad
	\subfloat{\includegraphics[height=1.3cm,width=1.3cm]{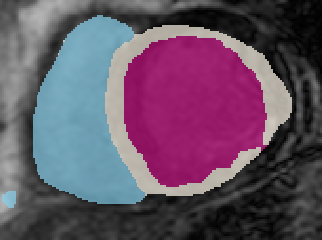}}\renewcommand{\thesubfigure}{} \quad
	\vspace{-0.3cm}
	
    U+A+D+T
	\subfloat{\includegraphics[height=1.3cm,width=1.3cm]{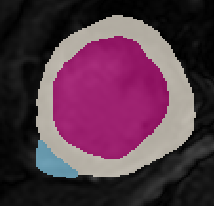}}\renewcommand{\thesubfigure}{} \quad
	\subfloat{\includegraphics[height=1.3cm,width=1.3cm]{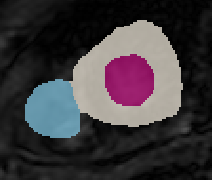}}\renewcommand{\thesubfigure}{} \quad
	\subfloat{\includegraphics[height=1.3cm,width=1.3cm]{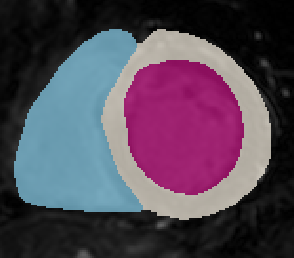}}\renewcommand{\thesubfigure}{} \quad
	\subfloat{\includegraphics[height=1.3cm,width=1.3cm]{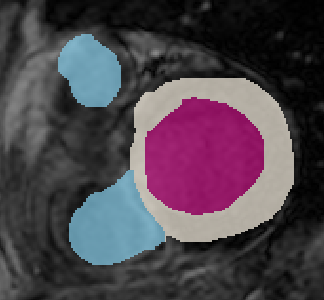}}\renewcommand{\thesubfigure}{} \quad
	\subfloat{\includegraphics[height=1.3cm,width=1.3cm]{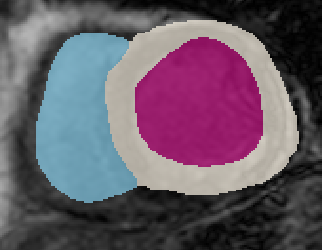}}\renewcommand{\thesubfigure}{} \quad
	\vspace{-0.1cm}
	
    \caption{The results of the segmentation. Rows from top to bottom: original images(O), dilated residual networks (U+D), adversarial model(U+A+D), adversarial model with Dice coefficient loss and transfer loss (U+A+D+T) (best viewed in color).}
	\label{fig:vis}
\end{figure}
\subsubsection{Visualisation.} Figure \ref{fig:vis} shows some predicted masks of the LGE slices of four patients. It could be seen that adversarial learning improves the results of only using the dilated residual network, and the cross-modality transfer further refine the segmentation masks, especially for the left ventricles. Those observations are consistent with the results shown in Tables \ref{tab:result}-\ref{tab:result_sub2}.

\section{Conclusions}
We propose an automated method for heart segmentation based on multi-modality MRI images, which is trained in an adversarial manner. Specifically, our architecture consists of two modules, a multi-channel mask generator and a discriminator. In particular, we further introduce a domain-transfer loss function to leverage the information across different modalities for the same patients. Results show that such an idea is effective, and the overall framework performs better than the baseline methods.

%
%
%
%

\end{document}